\documentclass[pra,aps,preprint,showpacs,superscriptaddress]{revtex4-1}
\usepackage{amssymb}
\usepackage{amsmath}
\usepackage{graphicx}
\usepackage{bm}
\usepackage{booktabs}
\usepackage{dcolumn}
\newcommand{\ket}[1]{\left| #1 \right>}
\newcommand{\bra}[1]{\left< #1 \right|}
\newcommand{\avg}[1]{\langle #1 \rangle}
\begin{document}
\title{Cyclotron Dynamics of a Kondo Singlet in a Spin-Orbit-Coupled Alkaline-Earth Atomic Gas}

\affiliation{Key Laboratory for Quantum Optics, Shanghai Institute of Optics and Fine Mechanics, Chinese Academy of Sciences, Shanghai 201800, China}
\affiliation{State Key Laboratory of High Field Laser Physics, Shanghai Institute of Optics and Fine Mechanics, Chinese Academy of Sciences, Shanghai 201800, China}
\affiliation{University of Chinese Academy of Sciences, Beijing 100049, China}

\author{Bo-Nan Jiang}
\affiliation{Key Laboratory for Quantum Optics, Shanghai Institute of Optics and Fine Mechanics, Chinese Academy of Sciences, Shanghai 201800, China}
\affiliation{University of Chinese Academy of Sciences, Beijing 100049, China}

\author{Hao Lv}
\affiliation{Key Laboratory for Quantum Optics, Shanghai Institute of Optics and Fine Mechanics, Chinese Academy of Sciences, Shanghai 201800, China}
\affiliation{University of Chinese Academy of Sciences, Beijing 100049, China}

\author{Wen-Li Wang}
\affiliation{Key Laboratory for Quantum Optics, Shanghai Institute of Optics and Fine Mechanics, Chinese Academy of Sciences, Shanghai 201800, China}
\affiliation{State Key Laboratory of Precision Spectroscopy, East China Normal University, Shanghai 200241, China}

\author{Juan Du}
\email{dujuan@mail.siom.ac.cn}
\affiliation{State Key Laboratory of High Field Laser Physics, Shanghai Institute of Optics and Fine Mechanics, Chinese Academy of Sciences, Shanghai 201800, China}

\author{Jun Qian}
\email{jqian@mail.siom.ac.cn}
\affiliation{Key Laboratory for Quantum Optics, Shanghai Institute of Optics and Fine Mechanics, Chinese Academy of Sciences, Shanghai 201800, China}

\author{Yu-Zhu Wang}
\email{yzwang@mail.shcnc.ac.cn}
\affiliation{Key Laboratory for Quantum Optics, Shanghai Institute of Optics and Fine Mechanics, Chinese Academy of Sciences, Shanghai 201800, China}

\date{\today}

\begin{abstract}
We propose a scheme to investigate the interplay between Kondo-exchange interaction and quantum spin Hall effect with ultracold fermionic alkaline-earth atoms trapped in two-dimensional optical lattices using ultracold collision and laser-assisted tunneling. In the strong Kondo-coupling regime, though the loop trajectory of the mobile atom disappears, collective dynamics of an atom pair in two clock states can exhibit an unexpected spin-dependent cyclotron orbit in a plaquette, realizing the quantum spin Hall effect of the Kondo singlet. We demonstrate that the collective cyclotron dynamics of the spin-zero Kondo singlet is governed by an effective Harper-Hofstadter model in addition to second-order diagonal tunneling.
\end{abstract}
\maketitle

\section{Introduction}
In the milestone work by Jun Kondo, the resistance minimum at nonzero temperature in a metal was explained by the scattering of conduction electrons by localized magnetic impurities \cite{Kondo}. Since that, Kondo effect has been considered as a primary mechanism in heavy-fermion systems \cite{Hewson}. After the description of this effect in the ground state of the Kondo lattice model (KLM) was formulated, the formation of Kondo singlet between mobile electron and localized spin is found to perform a key function \cite{Vonsovsky,Zener,Ueda1997,Cox}. Thanks to the rapid development of laser cooling and trapping techniques, neutral atoms have been considered as an ideal platform for simulating complicated phenomena in condensed matter physics \cite{Bloch2012}. For example, quantum mixtures of alkali atoms are expected to display the characteristics of Kondo-correlated state \cite{Bauer2013} and multi-orbital effect \cite{Nishida2013}. In particular, fermionic alkaline-earth atoms (AEAs) in optical lattices possess unique properties for operating optical atomic clocks with unprecedented precision~\cite{Derevianko,Ye2014nat} and provide novel insights into the physics of strongly correlated transition-metal oxides and heavy-fermion materials as quantum simulators \cite{Gorshkov,Rey2010,Rey2012,Kyoto2011,Kyoto2012}, where the interorbital spin-exchange interaction between $^1$S$_0$ and $^3$P$_0$ clock-state atoms plays an essential role \cite{Ye2014,Bloch2014a,Inguscio2014}.

Active interest has recently been focused on operating neutral atoms as charged particles by engineering artificial gauge potentials \cite{Dalibard,Spielman2009a,Spielman2009b,Zhang2012,Zwierlein2012}. As a significant breakthrough, spin Hall effect has been observed in a Bose-Einstein condensate, leading to the realization of a cold-atom spin transistor \cite{Spielman2013}, where the internal and external states of atoms are coupled by a two-photon Raman process. For trapped atoms in optical lattices, tunable gauge potentials have been implemented by periodic driving \cite{Sengstock} and ``Zeeman lattice'' \cite{Spielman2012} techniques. Furthermore, the Harper-Hofstadter Hamiltonian \cite{Harper,Hofstadter} has been experimentally realized with ultracold atoms in optical lattices \cite{Bloch2013,Ketterle2013} using laser-assisted tunneling (LAT) \cite{Kolovsky}, leading to a spin-dependent magnetic field and realizing the quantum spin Hall (QSH) effect \cite{Kennedy2013,Bloch2014b}. Moreover, since the interplay between interaction and QSH effect leads to compelling novel physics, both cyclotron motion of repulsive bosons in optical lattices \cite{Li2013} and QSH insulators with Kondo-exchange interaction \cite{Vonsovsky,Zener} in graphene or quantum wells \cite{Wu2006,Maciejko2009,Maciejko2011,Qi2011,Goth2013,Maciejko2012,Si2013} have been extensively discussed.

Motivated by these works, we present a proposal to study the interplay between Kondo-exchange interaction and QSH effect with ultracold fermionic AEAs trapped in two-dimensional optical lattices using ultracold collision and LAT. The underlying physics is probed by the cyclotron dynamics of a mobile atom and a Kondo singlet in a plaquette. We find that the cyclotron orbit of mobile atom is damped with nonzero Kondo coupling. More interestingly, at strong Kondo coupling, we predict a striking and unexpected phenomenon that the spin-zero Kondo singlet can behave as a spin-half particle exposed to a perpendicular spin-dependent magnetic flux. We demonstrate that the spin state of localized background atoms and second-order diagonal tunneling have nontrivial effects on the spin-dependent cyclotron dynamics of the Kondo singlet.

\section{Model}
\begin{figure}[tbp]
\includegraphics[width=0.8\linewidth]{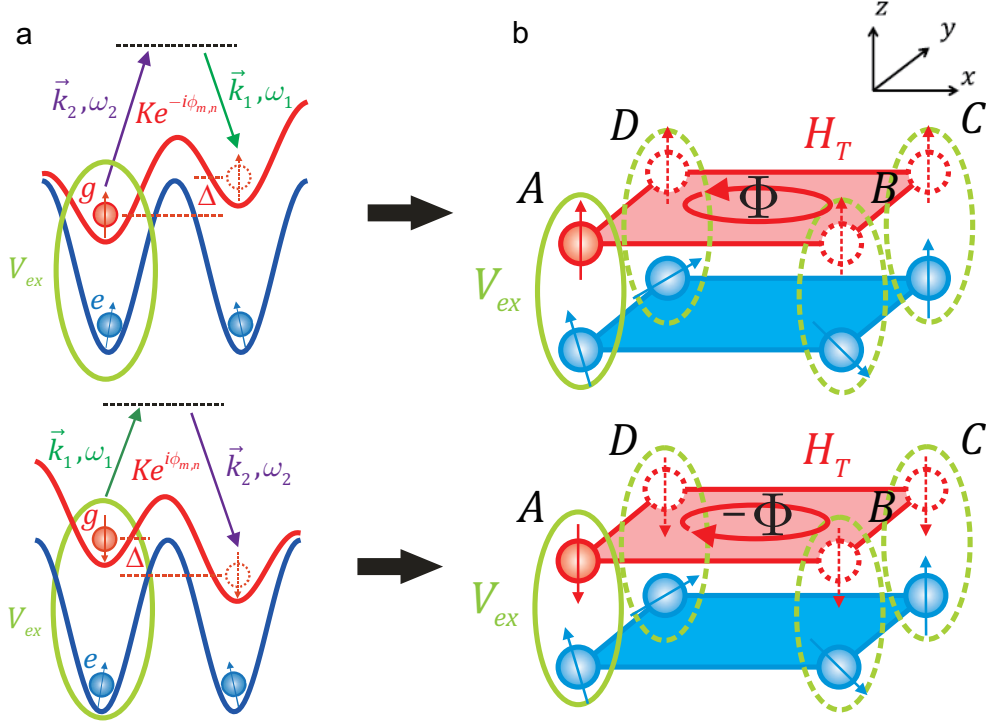}
\caption{(Color online) (a) Laser-assisted tunneling and Kondo coupling for $\ket{\uparrow}$ (top) and $\ket{\downarrow}$ (bottom) $g$ atoms. Though normal tunneling in the $x$ direction is suppressed by a magnetic field gradient $\Delta$ for two $g$ atom spin states, two Raman beams with frequency detuning $\delta\omega=\omega_2-\omega_1=\Delta$ and momentum transfer $\delta\vec{k}=\vec{k}_2-\vec{k}_1$ restore the resonant tunneling with a spin-dependent complex amplitude $K e^{\pm i \phi_{m,n}}$. (b) Schematic picture of effective magnetic flux in a plaquette with four sites $A, B, C,$ and $D$, where the cyclotron orbits of the $g$ atom that realize the spin-dependent effective magnetic flux $\pm\Phi$ are coupled to a plaquette of localized $e$ spins via onsite interorbital spin-exchange interaction (Kondo coupling) $V_{ex}$. $H_{T}$ denotes the Harper Hamiltonian.}
\label{fig:model}
\end{figure}

We consider fermionic AEAs in $^1$S$_0$ ($g$) and $^3$P$_0$ ($e$) clock states independently trapped in two-dimensional optical lattices of the same periodicity \cite{Daley} as sketched in Fig. \ref{fig:model}, where $\ket{\uparrow/\downarrow}$ denotes nuclear spin states $\ket{I,\pm m_I}$ \cite{Gorshkov,Rey2010,Rey2012}. The $\ket{\uparrow/\downarrow}$ $g$ atom (conduction electron) receives opposite frequency detuning $\delta\omega=\pm(\omega_2-\omega_1)=\pm\Delta$ and momentum transfer $\delta\vec{k}=\pm(\vec{k}_2-\vec{k}_1)$ from two far-detuned Raman beams and tunnels spin-dependently in the optical lattice tilted by a magnetic field gradient $\Delta$ along $x$ \cite{Kennedy2013}. In addition, the Mott insulator background of $e$ atoms (localized spins) interacts with the $g$ atom via onsite interorbital spin-exchange interaction $V_{ex} \propto (a^{-}_{eg} - a^{+}_{eg})\int d^3\mathbf{r}w^2_g(\mathbf{r}) w^2_e(\mathbf{r})$, where $a^{\pm}_{eg}$ is the scattering lengths for two atoms in $\ket{\pm}=\frac{1}{\sqrt{2}}(\ket{ge}\pm\ket{eg})$ and $w_{g(e)}(\mathbf{r})$ denotes the wavefunction for the $g$ ($e$) atom \cite{Gorshkov,Inguscio2014,Ketterle2013}. By averaging out rapidly oscillating terms in a rotating frame \cite{Bloch2013,Ketterle2013}, we obtain the time-independent Harper-Kondo Hamiltonian (see Appendix A):
\begin{eqnarray}
H &=& H_T+H_K \nonumber\\
    &=& - \sum_{m,n,\alpha}( K e^{-i \phi_{m,n,\alpha}} c^{\dag}_{m+1,n,\alpha} c_{m,n,\alpha}
	  + J c^{\dag}_{m,n+1,\alpha} c_{m,n,\alpha} + h.c.) \nonumber\\
    &&+ V_{ex} \sum_{m,n} \mathbf{s}^{c}_{m,n} \cdot \mathbf{S}^{f}_{m,n},
\label{eq:hamiltonian}
\end{eqnarray}
where $\mathbf{s}^{c}_{m,n} = \frac{1}{2} \sum\limits_{\alpha, \beta} c^{\dag}_{m,n,\alpha} \vec{\sigma}_{\alpha \beta} c_{m,n,\beta}$ and $\mathbf{S}^{f}_{m,n} = \frac{1}{2} \sum\limits_{\alpha, \beta} f^{\dag}_{m,n,\alpha} \vec{\sigma}_{\alpha \beta} f_{m,n,\beta}$ denote the spin operators of the $g$ and $e$ atoms at (m,n) ($\alpha, \beta \in \{ \uparrow, \downarrow \}$ and $\vec{\sigma}$ is the vector of Pauli matrices). $c^{\dag}_{m,n,\alpha}$ ($f^{\dag}_{m,n, \alpha}$) creates a $g$ ($e$) atom in the spin state $\ket{\alpha}$ at (m,n). $K e^{-i \phi_{m,n,\alpha}}$ is the spin-dependent complex tunneling amplitude of the mobile atom in the $x$ direction induced by LAT, while $J$ is the real tunneling amplitude of the mobile atom in the $y$ direction. The first term in Eq. (\ref{eq:hamiltonian}), i.e. the Harper Hamiltonian $H_{T}$ \cite{Harper,Hofstadter}, describes the $g$ atom in a spin-dependent magnetic field arising from spatially varying phase $\phi_{m,n,\uparrow/\downarrow}=\pm\phi_{m,n}=\pm\pi/2(m+n)$, realizing an effective magnetic flux $\Phi=\pm\pi/2$ for $\ket{\uparrow/\downarrow}$ $g$ atom and leading to spin-orbit coupling and QSH effect \cite{Kennedy2013}. The last term is the Kondo-exchange interaction $H_K$, which includes the spin-flip process that scatters the $g$ atom from $\ket{\uparrow/\downarrow}$ to $\ket{\downarrow/\uparrow}$ by exchanging the spin with the $\ket{\downarrow/\uparrow}$ $e$ atom at the same site.

\section{Damped orbital dynamics of the mobile atom}
\begin{figure}[tbp]
\includegraphics[width=0.5\textwidth]{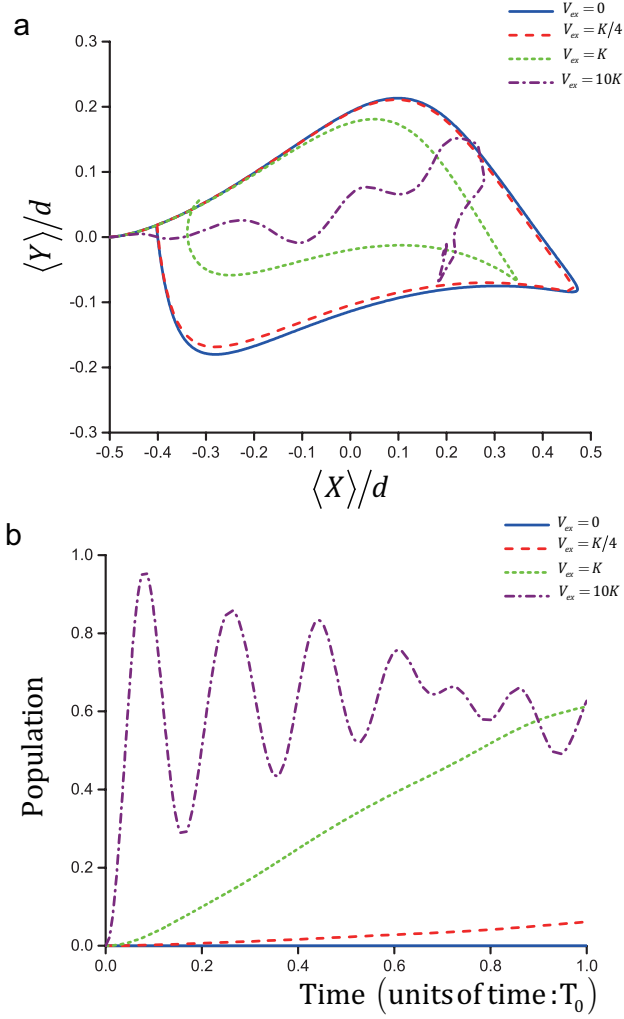}
\caption{(Color online) The cyclotron trajectory of the $g$ atom (a) and the corresponding time evolution of the spin-defect population $\sum_{m,n,m',n'}|\gamma^{m',n'}_{m,n}(t)|^2$ (b) at different Kondo-coupling strengths $V_{ex}=$ 0 (blue solid), $K/4$ (red dashed), $K$ (green dotted), and $10K$ (purple dash-dotted) in the period of $T_0$.}
\label{fig:single}
\end{figure}

\begin{figure}[tbp]
\includegraphics[width=0.5\textwidth]{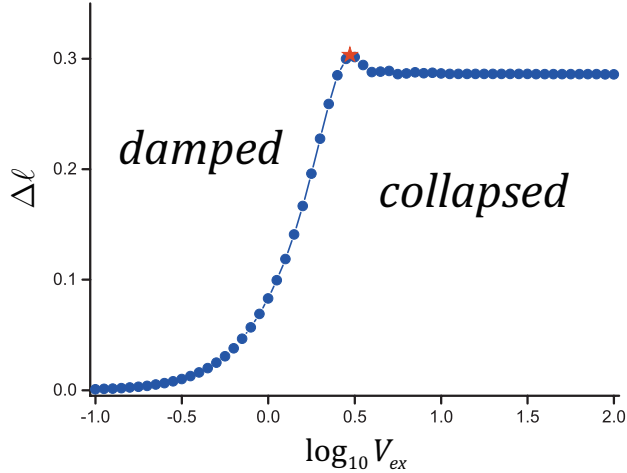}
\caption{(Color online) Time-averaged interaction-induced deviation $\Delta\ell$ from the cyclotron orbit of noninteracting mobile atom in the period of $T_0$ as a function of Kondo coupling $V_{ex}$. And the transition from the damped cyclotron orbit to the collapsed trajectory happens at $V_{ex}=J+K$, which is marked by the red star.}
\label{fig:transition}
\end{figure}

We revisit the cyclotron dynamics of the single noninteracting $\ket{\downarrow}$ mobile atom in Ref. \cite{Bloch2013} when Kondo-exchange interaction couples the mobile atom to a plaquette of polarized $\ket{\uparrow}$ $e$ atoms. The motion of the mobile atom here can be derived in the Hilbert space with quantum number $S_{tot}=S^z_{tot}=\frac{3}{2}$ from the state $\ket{\Psi(t)}=\sum\limits_{m,n}\left(\gamma_{m,n}(t)c^{\dag}_{m,n,\downarrow} +\sum\limits_{m',n'}\gamma^{m',n'}_{m,n}(t)c^{\dag}_{m,n,\uparrow}S^{f-}_{m',n'}\right)\ket{FM}$, where $\ket{FM}$ denotes the polarized $\ket{\uparrow}$ $e$ atoms at four sites $A, B, C,$ and $D$ associated with the $g$ atom vacuum \cite{Mattisa,Mattisb,Katsnelsona,Katsnelsonb,Ueda1991a,Ueda1991b}, and $S^{f-}_{m,n}=f^{\dag}_{m,n,\downarrow}f_{m,n,\uparrow}$. $\gamma_{m,n}$ is the probability amplitude of $\ket{\downarrow}$ $g$ atom, and $\gamma^{m',n'}_{m,n}$ that is relevant to the spin-flip process between $\ket{\downarrow}$ $g$ atom and $\ket{\uparrow}$ $e$ atom is the probability amplitude of $\ket{\downarrow}$ $e$ spin defect in a background of $\ket{\uparrow}$ $e$ atoms. We consider the time evolution of the system initially prepared in $\ket{\Psi_0}=\frac{1}{\sqrt{2}}\left(c^{\dag}_{A,\downarrow}+c^{\dag}_{D,\downarrow}\right)\ket{FM}$ with $K=2\pi\hbar\times(0.27\mathrm{kHz})$ and $J=2\pi\hbar\times(0.53\mathrm{kHz})$ in a period of $T_0=2.3\mathrm{ms}$ as Ref. \cite{Bloch2013}, and during $T_0$, the $\ket{\downarrow}$ $g$ atom without Kondo coupling can reproduce the loop trajectory in Ref. \cite{Bloch2013}. By solving the Schr\"{o}dinger equation with Eq. (\ref{eq:hamiltonian}), we obtain the mean position of the $g$ atom along $x$ and $y$, $\langle X \rangle = (N_B + N_C - N_A - N_D) d /2$ and $\langle Y \rangle = (N_C + N_D - N_A - N_B) d /2$, with the onsite population of the $g$ atom $N_{m,n}$ ($(m,n)\in A,B,C,D$) and lattice constant $d$. The trajectories of the mobile atom for different strengths of the Kondo coupling $V_{ex}$ are shown in Fig. \ref{fig:single}(a). At weak Kondo coupling or around quantum critical point ($V_{ex} = K/4, K$), the cyclotron orbit gradually shrinks; and at strong Kondo coupling ($V_{ex} = 10 K$), the loop trajectory is totally destroyed, and the atom moves irregularly. Different from cyclotron motion damping induced by repulsive interaction in Ref. \cite{Li2013}, the damping of the cyclotron orbit of the $g$ atom is attributed to the spin-flip process induced by Kondo-exchange interaction. As a consequence of spin flips, the $\ket{\downarrow}$ $g$ atom evolves into a superposition of $\ket{\uparrow}$ and $\ket{\downarrow}$, and since the $\ket{\uparrow}$ and $\ket{\downarrow}$ $g$ atoms manifest opposite chiralities of motion in the spin-dependent magnetic field, the cyclotron orbit shall be evidently damped \cite{Bloch2013}. Because the spin exchange between $\ket{\downarrow}$ $g$ atom and $\ket{\uparrow}$ $e$ atom creates a $\ket{\downarrow}$ spin defect in polarized $\ket{\uparrow}$ $e$ atoms, the time evolution of the spin-defect population $\sum_{m,n,m',n'}|\gamma^{m',n'}_{m,n}(t)|^2$ in Fig. \ref{fig:single}(b) monitors the spin-flip event during the motion of the $g$ atom. At nontrivial Kondo coupling, the fractional spin defect population confirms the occurrence of spin flips and the superposition of $\ket{\uparrow}$ and $\ket{\downarrow}$ $g$ atoms; from $V_{ex} = K/4$ to $K$, the spin defect population remarkably increases in the same period as a result of strengthening the spin-flip process; and the intriguing oscillation at $V_{ex} = 10K$ can be considered as a signature of the spin-current vortex in Ref. \cite{Wu2006}, which is related to the formation of composite objects under strong spin flips \cite{Hewson,Ueda1991b}. Furthermore, we define the time-averaged interaction-induced deviation from the cyclotron orbit of noninteracting $\ket{\downarrow}$ mobile atom in the period of $T_0$ as $\Delta\ell=\frac{1}{T_0}\int^{T_0}_{0}\mathrm{d}t\sqrt{[\Delta X(t)]^2+[\Delta Y(t)]^2}$, where $\Delta X(t)=\avg{X(t)}_{V_{ex}}-\avg{X(t)}_0$ and $\Delta Y(t)=\avg{Y(t)}_{V_{ex}}-\avg{Y(t)}_0$. $\avg{X(t)}_{V_{ex}}$ and $\avg{Y(t)}_{V_{ex}}$ are the mean position of the $g$ atom along $x$ and $y$ at time $t$ and Kondo coupling strength $V_{ex}$. And a more systematic study of the effect of Kondo coupling on the cyclotron orbit is launched by monitoring $\Delta\ell$ as a function of $V_{ex}$ in Fig. \ref{fig:transition}. At weak Kondo coupling or around quantum critical point, the deviation $\Delta\ell$ increases with the strength of Kondo-exchange interaction, indicating that the cyclotron orbit is being further damped, which is consistent with the trajectories derived in the same parameter regime in Fig. \ref{fig:single}(a). The damped cyclotron orbit of the mobile atom begins to collapse when the Kondo coupling $V_{ex}$ overwhelms the kinetic energy of the noninteracting mobile atom in the plaquette $J+K$ (red star), where the collapsed trajectory manifests itself by the constant deviation $\Delta\ell$ with varying $V_{ex}$ at strong Kondo coupling.

\section{Collective cyclotron dynamics of the Kondo singlet}
\begin{figure}[tbp]
\includegraphics[width=0.5\textwidth]{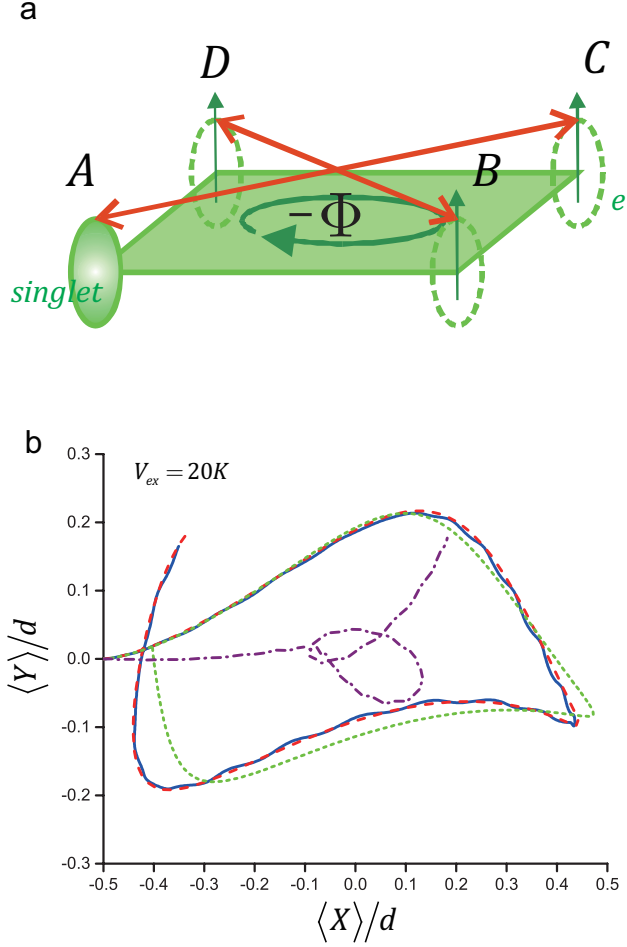}
\caption{(Color online) (a) Schematic illustration of the effective magnetic field and diagonal tunneling for the Kondo singlet in polarized $\ket{\uparrow}$ $e$ atoms. (b) The cyclotron trajectories of the Kondo singlet in the period of $2T_0$ are obtained from $H$ (blue solid), $H_{eff}$ (red dashed) and $H^{(s)}_T$ (green dotted). If $e$ spins become nonparallel, for example, by flipping the $e$ spin at B, the cyclotron trajectory disappears (purple dash-dotted).}
\label{fig:singlet}
\end{figure}

The damping of the cyclotron orbit above shows that the QSH effect of the $g$ atom \cite{Bloch2013,Kennedy2013} is not robust to Kondo-exchange interaction \cite{Maciejko2011,Qi2011}. At strong Kondo coupling, since the $g$ and $e$ atoms at the same site tend to form a Kondo singlet \cite{Hewson,Ueda1991b}, we naturally investigate the behavior of a composite object in the synthetic magnetic field. First we define the creation operators in a background of $\ket{\uparrow}$ $e$ atoms for the Kondo singlet $s^{\dagger}_{m,n}=\frac{1}{\sqrt{2}}(c^{\dagger}_{m,n,\uparrow}f^{\dagger}_{m,n,\downarrow}-c^{\dagger}_{m,n,\downarrow}f^{\dagger}_{m,n,\uparrow})f_{m,n,\uparrow}$ and triplets $t^{0\dagger}_{m,n}=\frac{1}{\sqrt{2}}(c^{\dagger}_{m,n,\uparrow}f^{\dagger}_{m,n,\downarrow}+c^{\dagger}_{m,n,\downarrow}f^{\dagger}_{m,n,\uparrow})f_{m,n,\uparrow}$ and $t^{1\dagger}_{m,n}=S^{f-}_{m',n'}c^{\dagger}_{m,n,\uparrow}f^{\dagger}_{m,n,\uparrow}f_{m,n,\uparrow}$, where $S^{f-}_{m',n'}$ with $m',n'\neq m,n$ is a restriction by the requirement of $S_{tot}=S^z_{tot}=\frac{3}{2}$ in the Hilbert space. The physical interpretation of this definition is that the creation of a composite object is accompanied by the annihilation of the $e$ atom at the same site. Then, in the singlet-triplet representation, the Harper-Kondo Hamiltonian can be partitioned as \cite{Auerbach}
\begin{equation}
\label{eq:hmatrix}
H = \left( \begin{array}{cccc}
\mathcal{P}_s H_T \mathcal{P}_s -\frac{3 V_{ex}}{4} & \mathcal{P}_s H_T \mathcal{P}_t \\
\mathcal{P}_t H_T \mathcal{P}_s & \mathcal{P}_t H_T \mathcal{P}_t + \frac{V_{ex}}{4}
\end{array} \right).
\end{equation}
$\mathcal{P}_{s,t}$ are the projectors onto the subspaces $\{\ket{s}_{m,n}=s^{\dagger}_{m,n}\ket{FM}\}$ and $\{\ket{t^{0,1}}_{m,n}=t^{0,1\dagger}_{m,n}\ket{FM}\}$ whose Kondo-exchange energy are $-\frac{3 V_{ex}}{4}$ and $\frac{V_{ex}}{4}$, respectively. Considering that the coupling between $\{\ket{s}_{m,n}\}$ and $\{\ket{t^{0,1}}_{m,n}\}$ is almost energetically forbidden by the large Kondo-exchange energy gap $V_{ex}\gg J,K$, the behavior of the Kondo singlet and triplet can be considered to be approximately independent. And the orientation of the magnetic field can be inferred from $\pm\phi_{m,n}$ of the nearest-neighbor tunneling matrix element $-\frac{K}{2} e^{\pm i \phi_{m,n}}$ in $\{\ket{s}_{m,n}\}$ and $\{\ket{t^{0,1}}_{m,n}\}$ \cite{Kennedy2013}. In the high-energy triplet subspace, opposite magnetic fields simultaneously work during the tunneling of the triplets: $\ket{t^{0,1}}_{m,n}\rightarrow\ket{t^{0,1}}_{m+1,n}$ with $\phi_{m,n}$ and $\ket{t^{0,1}}_{m,n}\rightarrow\ket{t^{1,0}}_{m+1,n}$ with $-\phi_{m,n}$. Consequently, the triplets' movement neither realizes effective magnetic flux nor manifests chirality \cite{Bloch2013}. However, in the low-energy singlet subspace, the Kondo singlet in a background of $\ket{\uparrow}$ $e$ atoms experiences a unidirectional magnetic field arising from $\phi_{m,n}=\pi/2(m+n)$. The effective Hamiltonian in the low-energy subspace $\{\ket{s}_{m,n}\}$ is explicitly given by \cite{Auerbach} (see Appendix B for more details)
\begin{eqnarray}
\label{eq:heff}
	H_{eff} &=& \mathcal{P}_s H_{T} \mathcal{P}_s +
		\mathcal{P}_s H_T \frac{1}{\mathcal{P}_t [ -\frac{3}{4} V_{ex}-(H_T + H_{K}) ]  \mathcal{P}_t} H_T \mathcal{P}_s \nonumber \\
	&=& - \sum_{m,n} \left(\frac{K}{2} e^{i \phi_{m,n}} s^{\dag}_{m+1,n} s_{m,n} + \frac{J}{2} s^{\dag}_{m,n+1} s_{m,n} + h.c.\right) \nonumber \\
    & & - \sum_{m,n} \{J_d \left[ e^{i (\phi_{m,n} + \phi_{m+1,n})} + e^{i (\phi_{m,n} + \phi_{m,n+1})} \right] s^{\dag}_{m+1,n+1} s_{m,n} + h.c.\} \nonumber \\
	&=& H^{(s)}_T + H^{(s)}_d,
\end{eqnarray}
where $J_d = \frac{J K}{4 V_{ex}}$ and trivial constants are neglected. The first term in Eq. (\ref{eq:heff}) reconstructs the Harper Hamiltonian $H^{(s)}_T$ of Kondo singlet, and the factor $1/2$ originates from the fact that the effective mass of Kondo singlet is twice the mass of the g atom \cite{Ueda1991b}. The last term $H^{(s)}_d$ describes the second-order diagonal tunneling of Kondo singlet from $A$ to $C$ or $D$ to $B$. Then, $e^{i (\phi_{m,n} + \phi_{m+1,n})} + e^{i (\phi_{m,n} + \phi_{m,n+1})}$ reflects the interference between two second-order hopping paths, which reduces to $1+e^{i \frac{\pi}{2}}$ for the paths $A \rightarrow B \rightarrow C$ and $A \rightarrow D \rightarrow C$. Following the procedure above, the time-reversal counterpart of Eq. (\ref{eq:heff}) with $\phi_{m,n} \rightarrow -\phi_{m,n}$ can also be obtained in a background of $\ket{\downarrow}$ $e$ atoms. This means that the Kondo singlet experiences a $e$-spin-dependent magnetic field as a spin-half particle, and the opposite chiralities of the Kondo singlet's cyclotron orbit shall be locked to polarized $\ket{\uparrow}$ and $\ket{\downarrow}$ $e$ atoms respectively \cite{Bloch2013,Kennedy2013}, thus realizing the QSH effect of the Kondo singlet as a result. From the aforementioned demonstrations where the QSH effect of the $g$ atom is destroyed at strong Kondo coupling and the spin-dependent magnetic field on the Kondo singlet is irrelevant to the spin state of the $g$ atom, one can realize that the QSH effect of the Kondo singlet revealed in our scheme is nontrivial.

The Harper Hamiltonian and diagonal tunneling for the Kondo singlet in the background of $\ket{\uparrow}$ $e$ atoms are illustrated in Fig. \ref{fig:singlet}(a). We consider the cyclotron motion of the Kondo singlet in the initial state $\frac{1}{\sqrt{2}} ( \ket{s}_A + \ket{s}_D)$ during the period of $2T_0$ at $V_{ex}=20K$. In Fig. \ref{fig:singlet}(b), different trajectories of the Kondo singlet are calculated by $H$, $H_{eff}$ and $H^{(s)}_T$. Clearly, the effective Hamiltonian $H_{eff}$ at strong Kondo coupling resembles the cyclotron dynamics obtained by solving the Schr\"{o}dinger equation with the Harper-Kondo Hamiltonian $H$ and accurately describes the cyclotron orbit of the Kondo singlet in the plaquette. Furthermore, comparing the cyclotron trajectories obtained by $H_{eff}$ and $H^{(s)}_T$, we find an essential role of $H^{(s)}_d$ in the movement of Kondo singlet. The singlet travels a longer distance under $H_{eff}$ than that predicted by $H^{(s)}_T$, because $H^{(s)}_d$ speeds the singlet up around the plaquette. Moreover, the QSH effect of the Kondo singlet is not robust to interactions that destroy the parallel alignment of $e$ spins. To verify this statement, we flip one arbitrary spin of the polarized $e$ atoms in the initial state, for example, $S^{f-}_B\frac{1}{\sqrt{2}}(\ket{s}_A+\ket{s}_D)$, and find that the cyclotron orbital motion of the Kondo singlet disappears. The result indicates the competition between the QSH effect of the Kondo singlet and the interaction-induced nonparallel state of $e$ spins \cite{Si2013}.

\section{Experimental considerations}
Now we address practical considerations of observing the cyclotron trajectory of the Kondo singlet. Here we only consider ultracold $^{171} $Yb$ $ atoms in the optical lattice. The Raman lasers have to be detuned from the resonant frequency of 399 nm of the $^1$S$_0$$-$$^1$P$_1$ transition. For $^1$S$_0$ atoms trapped in a lattice ($\lambda=532$ nm) with the depth $V_0 = 15 E_{R}$, we have tunneling rate $t_x=31$ Hz and band gap $\omega=26$ kHz. Given Land\'{e} factor $g_I = 5.4 \times 10^{-4}$ and magnetic field gradient $B' \sim 2 \times 10^3$ mG/$\mu$m, energy tilting between nearest-neighbor sites is estimated to be $\Delta \approx 200$ Hz. Therefore, the requirement of $t_x \ll \Delta \ll \omega$ can be satisfied \cite{Miyake2013}. The spin-exchange strength $V_{ex}$ on the order of tens of kHz \cite{Ye2014,Bloch2014a,Inguscio2014} can be tuned by adjusting the lattice depth which changes the overlap of local atomic wavefunctions. Though the $^3$P$_0$ atoms can also feel the magnetic field gradient, their wavefunctions are only slightly affected because of the large lattice depth. The atom distribution can be detected by band-mapping technique in optical superlattices \cite{Bloch2013}. Key ingredients in our scheme can be realized by state-of-the-art techniques \cite{Bloch2013,Ketterle2013,Bloch2014b,Ye2014,Bloch2014a,Inguscio2014}.


\section{Conclusion}
In summary, we have studied the cyclotron dynamics of the mobile atom and the Kondo singlet in the plaquette with interorbital spin-exchange interaction and predicted a novel QSH effect of the spin-zero Kondo singlet in a background of polarized spins, demonstrating that the interplay between Kondo-exchange interaction and QSH effect constitutes more than competition \cite{Wu2006,Maciejko2009,Maciejko2011,Qi2011,Goth2013,Maciejko2012,Si2013} for a composite object. We show that the strong Kondo-exchange interaction destroys the loop trajectory of the mobile atom, but instead generates the cyclotron orbital motion for a composite object (Kondo singlet). For spin-orbit-coupled heavy fermions in optical square lattices, our work suggests the competition between the QSH effect of the Kondo singlet and the RKKY-induced Neel state \cite{Si2013,Lacroix1979}, which provides opportunities for novel quantum phases and critical points \cite{Si2013,Si2010a,Si2010b}. As such, we hope our work may pave a way for experiments on spin-orbit-coupled heavy-fermion systems with AEAs in the future.

\section*{Acknowledgments}
We thank Colin Kennedy, Hui Hu and Xia-Ji Liu for helpful discussions. This work was supported by National Foundation of Research Program (No. 2011CB921504) and NSFC (No. 11104292). J.D. thanks the support by 100 Talents Program of CAS. W.-L. W. thanks the support by Open Research Fund of State Key Laboratory of Precision Spectroscopy (East China Normal University) and NSFC (No. 11404353).

\appendix
\section{Harper-Kondo Hamiltonian with laser-assisted tunneling}
We extend the scheme of realizing an Abelian gauge field with spin-dependent laser-assisted tunneling \cite{Kennedy2013} to the system considered in the main text as follows: although normal tunneling in the $x$ direction is suppressed by a magnetic field gradient $\Delta$ for $g$ atoms with opposite magnetic moments, a driven Raman process with driving frequency $\omega$ and two-photon Rabi frequency $\Omega$ restores resonant tunneling. We average out rapidly oscillating terms in a rotating frame and obtain the time-independent Harper Hamiltonian of the $g$ atom:
\begin{equation}
\label{S1}
H_{T} = - \sum_{m,n,\alpha}( K e^{-i\phi_{m,n,\alpha}} c^{\dagger}_{m+1,n,\alpha} c_{m,n,\alpha} + J c^{\dagger}_{m,n+1,\alpha} c_{m,n,\alpha} + h.c.),
\end{equation}
where $\phi_{m,n,\uparrow/\downarrow}=\pm\phi_{m,n}$. Subsequently, we generalize the unitary operator in Ref. \cite{Miyake2013} to the spin degree of freedom
\begin{equation}
\label{S2}
U = \sum_{m,n,\alpha}e^{-i[m\omega t+\frac{\Omega}{\omega}\cos(\omega t-\phi_{m,n,\alpha})]}c^{\dagger}_{m,n,\alpha}c_{m,n,\alpha},
\end{equation}
and derive the expression of Kondo-exchange interaction in the above rotating frame at resonant tunneling
\begin{eqnarray}
\label{S3}
H_K &=& V_{ex} \sum_{m,n} U^{\dagger} s^{c,z}_{m,n} S^{f,z}_{m,n} U + \frac{1}{2T} \int^{\frac{T}{2}}_{-(\frac{T}{2})}\mathrm{d}tU^{\dagger}(s^{c+}_{m,n}S^{f-}_{m,n}+s^{c-}_{m,n}S^{f+}_{m,n})U \nonumber \\
&=& V_{ex} \sum_{m,n} s^{c,z}_{m,n} S^{f,z}_{m,n} + \left[ \frac{1}{2T} \int^{\frac{T}{2}}_{-\frac{T}{2}} e^{i\nu\omega t} \mathrm{d}t \sum_{\nu} J_{\nu} \left( \frac{2 \Omega}{\omega} \sin \phi_{m,n} \right)  s^{c+}_{m,n} S^{f-}_{m,n} + h.c. \right] \nonumber \\
&=& V_{ex} \sum_{m,n} s^{c,z}_{m,n} S^{f,z}_{m,n}  +  \frac{1}{2} J_0 \left( \frac{2\Omega}{\Delta}\sin\phi_{m,n} \right) (s^{c+}_{m,n}S^{f-}_{m,n}+s^{c-}_{m,n}S^{f+}_{m,n}),
\end{eqnarray}
where $J_{\nu}(x)$ are the Bessel functions of the first kind. The anisotropy of the term $J_0 (\frac{2\Omega}{\Delta} \sin \phi_{m,n})$ can often be relaxed. When $\frac{2\Omega}{\Delta}<\frac{1}{4}$, because $J_0(\frac{2\Omega}{\Delta}\sin\phi_{m,n})>0.985$, the anisotropy for an arbitrary phase $\phi_{m,n}$ can be reasonably ignored. Thus, in future experiments, by choosing a proper ratio between two-photon Rabi frequency and magnetic field gradient, the isotropic Kondo-exchange interaction can always be reached in the driven optical lattice:
\begin{equation}
\label{S4}
H_{K} \simeq V_{ex}\sum_{m,n} \mathbf{s}^c_{m,n} \cdot \mathbf{S}^f_{m,n}.
\end{equation}

\section{Projection to derive the effective Hamiltonian (Eq. (3) in the main text)}
In the main text, we obtain the effective Hamiltonian in the strong Kondo-coupling regime by the projection
\begin{equation}
\label{S5}
	H_{eff} = \mathcal{P}_s H_{T} \mathcal{P}_s +
		\mathcal{P}_s H_T \frac{1}{\mathcal{P}_t [ -\frac{3}{4} V_{ex}-(H_T + H_{K}) ]  \mathcal{P}_t} H_T \mathcal{P}_s. \nonumber
\end{equation}
Here, we divide $H_{eff}$ into two parts $H_1 = \mathcal{P}_s H_T \mathcal{P}_s$ and $H_2 = H_{eff} - H_1$, and the matrix elements of $H_{1,2}$ are explicitly given in the low-energy singlet subspace $\{ \ket{s}_{m,n} \}$ where $(m,n)$ represents plaquette sites $A, B, C,$ and $D$. $H_1$ includes only the nearest-neighbor tunneling term
\begin{equation}
_{m',n'} \bra{s} H_1 \ket{s}_{m,n} =
\begin{array}{c}
\begin{array}{ccccccccccc}
	\ket{s}_A & \ket{s}_B & & & \ket{s}_C  & & & & & \ket{s}_D &
\end{array} \\
\\
\left(\begin{array}{cccc}
	0 			 & -\frac{K}{2} &      0 	   & -\frac{J}{2} \\
    -\frac{K}{2} &     0 		& -\frac{J}{2} &      0       \\
    0 			 & -\frac{J}{2} &      0 	 &-\frac{K}{2}e^{i \pi/2} \\
    -\frac{J}{2} &     0        & -\frac{K}{2}e^{-i \pi/2} & 0
\end{array}\right),
\end{array}
\end{equation}
which gives $H^{(s)}_T$ in the main text. $H_2$ includes the onsite interaction, as well as the nearest-neighbor and diagonal tunneling terms. First, the onsite interaction term only contributes the trivial constant
\begin{equation}
   _{m,n} \bra{s} H_2 \ket{s}_{m,n} = -\frac{3}{4}\frac{K^2+J^2}{V_{ex}}.
\end{equation}
Second, the nearest-neighbor tunneling term can be safely neglected because
\begin{equation}
_{m+1,n} \bra{s} H_2 \ket{s}_{m,n} = _{m,n+1} \bra{s} H_2 \ket{s}_{m,n} = \mathit{O}\left( \frac{K}{V_{ex}} \right) \simeq 0.
\end{equation}
Third and last, we obtain the second-order diagonal tunneling as
\begin{equation}
   _{m+1, n+1} \bra{s} H_2 \ket{s}_{m,n} = -\frac{KJ}{4V_{ex}}(1+e^{i \pi/2}),
\end{equation}
where $1+e^{i \pi/2}$ originates from the interference between possible paths of diagonal tunneling from $A$ to $C$ or $D$ to $B$. For example, a Kondo singlet initially at site $A$ can pass through $B$ or $D$ to arrive at diagonal site $C$. In the tunneling path $A \rightarrow B \rightarrow C$, the singlet carries a zero phase because $\phi_A=0$ and $\phi_B=0$. However, in the other path $A \rightarrow D \rightarrow C$, the accumulated phase changes to $\phi_A + \phi_D = 0 + \frac{\pi}{2} = \frac{\pi}{2}$. As a result, the interference between two different tunneling paths contributes $e^{i 0}+e^{i \frac{\pi}{2}}=1+e^{i \frac{\pi}{2}}$. Finally, $H_2$ gives $H^{(s)}_d$ in the main text.

\end{document}